\documentclass[artice,aps,pra,showpacs,twocolumn,superscriptaddress]{revtex4}
\usepackage{graphicx,color}
\usepackage{amsmath,amsthm,amsfonts,amssymb,bm}
\usepackage{times}
\usepackage{epsf}
\usepackage[colorlinks={true}]{hyperref}
\usepackage[T1]{fontenc}
\usepackage[utf8]{inputenc}
\hypersetup{citecolor={blue}, filecolor={blue}, linkcolor={blue}, urlcolor={blue}}

\newcommand{\commentold}[1]{}
\DeclareMathSymbol{:}{\mathpunct}{operators}{"3A}
\usepackage[utf8]{inputenc}
\usepackage[english]{babel}
\usepackage{amsthm}
\theoremstyle{definition}

\begin{document}

\title{Controlling the entropic uncertainty lower bound in two-qubit systems under the decoherence}
\author{S. Haseli}
\email{soroush.haseli@uut.ac.ir}
\affiliation{Department of Physics, Urmia University of Technology, Urmia, Iran
}
\author{H. Dolatkhah}
\affiliation{Department of Physics, University of Kurdistan, P.O.Box 66177-15175, Sanandaj, Iran
} 
\author{S. Salimi}
\affiliation{Department of Physics, University of Kurdistan, P.O.Box 66177-15175, Sanandaj, Iran
}
\author{A. S. Khorashad}
\affiliation{Department of Physics, University of Kurdistan, P.O.Box 66177-15175, Sanandaj, Iran
}


\date{\today}

\begin{abstract}
The uncertainty principle is an inherent characteristic of quantum mechanics. This principle  can be formulated in various form. Fundamentally, this principle can be expressed in terms  of the standard deviation of the measured observables.  In quantum information theory the preferred mathematical quantity  to express the entropic uncertainty relation is the  Shannon's entropy. In this work, we consider the generalized entropic uncertainty relation in which there is an additional particle as a quantum memory. Alice measures on her particle $A$ and Bob, with memory particle $B$, predicts the Alice's measurement outcomes.  We study the effects of the environment on the entropic uncertainty lower bound in the presence of weak measurement and measurement reversal. The dynamical model that is intended in this work is as follows: First the weak measurement is performed, Second the decoherence affects on the system and at last the measurement reversal is performed on quantum system . Here we consider the generalized amplitude damping channel and depolarizing channel as environmental noises. We will show that in the presence of weak measurement and measurement reversal, despite the presence of environmental factors, the entropic uncertainty lower bound  dropped  to an optimal minimum value. In fact, weak measurement and measurement reversal enhance the quantum correlation between the subsystems $A$ and  $B$ thus the uncertainty of Bob about Alice's measurement outcomes reduces.   
\end{abstract}
\maketitle
\maketitle

\section{Introduction}
One of the inherent features of quantum mechanics is the uncertainty principle. This principle sets a limits on our ability to precise prediction of the measurement outcomes of two incompatible observables simultaneously. According to the Heisenberg uncertainty principle, it is not possible to measure the position and momentum  of a particle simultaneously  with high precision \cite{HEISENBERG}. In Ref.\cite{KENNARD}, based on the Heisenberg uncertainty principle,  Kennard  formulated the first uncertainty relation  of position $\hat{x}$ and momentum $\hat{p}$ as 
\begin{equation}
\Delta \hat{p} \Delta \hat{x} \geq \frac{\hslash}{2}.
\end{equation}
Robertson \cite{Robertson} and Schrodinger \cite{Schrodinger} have shown that for arbitrary pairs of incompatible observables $\hat{Q}$ and $\hat{R}$, the uncertainty relation is introduced in terms of the standard deviation as
\begin{equation}
\Delta \hat{Q} \Delta \hat{R} \geq \frac{1}{2} \vert \langle \psi \vert \left[ \hat{Q},\hat{R} \right] \vert \psi \rangle \vert,
\end{equation}
where $\Delta \hat{Q} = \sqrt{ \langle \psi \vert \hat{Q}^{2} \vert \psi \rangle - \langle \psi \vert \hat{Q} \vert \psi \rangle^{2}}$ and $\Delta \hat{R} = \sqrt{ \langle \psi \vert \hat{R}^{2} \vert \psi \rangle - \langle \psi \vert \hat{R} \vert \psi \rangle^{2}}$ are the standard deviations and $\left[\hat{Q},\hat{R} \right] = \hat{Q}~\hat{R}-\hat{R}~\hat{Q}$. The lower bound of this uncertainty relation is  depend  on the state of the system.  It  becomes trivial if the expectation value of the commutator $\left[\hat{Q},\hat{R} \right]$ on state $\vert \psi \rangle$ is zero. The uncertainty relation  is expressed in various forms. In quantum information theory the preferred mathematical quantity  to express the uncertainty relation is the Shannon's entropy \cite{Wehner}. With regard to the concept of Shannon's entropy, which indicates the amount of awareness about the measurement outcomes, it is quite logical to express the uncertainty relation in terms of Shannon entropy. In Ref. \cite{Kraus}, one of the most famous uncertainty relations  was presented by Kraus.  It was proved by Maassen and Uffink \cite{Maassen}
\begin{equation}\label{entropic}
H(\hat{Q})+H(\hat{R}) \geq \log_{2}\frac{1}{c},
\end{equation} 
where $H(\hat{O})=-\sum_{o} p_{o} \log_{2} p_{o}$ is the Shannon entropy of the measured observable $\hat{O} \in \lbrace \hat{Q}, \hat{R}\rbrace$, $p_o$ represents the possibility that the result of the measurement of the  observable $\hat{O}$   on the system  $\rho$ is $o$,  and $c=\max_{\lbrace i,j\rbrace} \vert \langle q_i \vert r_j \rangle \vert^{2}$ quantifies the ‘complementarity’ between the observables, where $ \vert q_i \rangle$ and $ \vert r_j \rangle$ are the eigenstates of the Hermitian observables $\hat{Q}$ and $\hat{R}$, respectively. In Ref. \cite{Berta}, Berta et al. examined the situation in which Bob has an extra particle as a quantum memory(particle $B$), which is entangled with the particle that is available for Alice(particle $A$). They showed that when Alice measures $\hat{Q}$ and $\hat{R}$, the  uncertainty of Bob ,that has access to memory particle, about the  Alice's measurement outcomes is bounded by
\begin{equation}\label{berta}
S(\hat{Q} \vert B)+ S(\hat{R} \vert B) \geq \log_2 \frac{1}{c} + S(A \vert B),
\end{equation} 
where $S(A \vert B)=S(AB)-S(B)$ is the conditional Von Neumann entropy, $S(\rho)=-tr(\rho \log_2 \rho)$denotes the Von Neumann entropy and $S(\hat{O} \vert B)=S(\rho^{\hat{O}B})-S(\rho^{B})$, $O \in \lbrace \hat{Q},\hat{R} \rbrace$ shows the conditional Von Neumann entropies of the post measurement states
\begin{equation}
\rho^{\hat{O}B}=\sum_{i} (\vert o_i \rangle \langle o_i \vert \otimes \mathcal{I})\rho^{AB} (\vert o_i \rangle \langle o_i \vert\otimes \mathcal{I}),
\end{equation}
where $\lbrace \vert o_i \rangle \rbrace$’s are the eigenstates of the observable $O$, and $\hat{\mathcal{I}}$ is the identity operator. Berta's entropic uncertainty relation has the vast applications in the field of quantum information such as  witnessing entanglement and cryptographic security\cite{Berta,Prevedel,Li}. Note that, much efforts have been made to improve Betta's entropic uncertainty relation \cite{Pati,Pramanik,Coles,Liu,Zhang,Pramanik1,Adabi,Adabi1,Dolatkhah,Jin-Long}.  In the real world, quantum systems interact with their surroundings, thus investigation of open quantum systems from various  perspectives has been subject of intense research in recent years. However, in
realistic quantum world, entanglement is inevitably affected by the interaction between the system and its environment, which leads to degradation. Given the importance of the entanglement between particle $A$ and $B$ in Berta's uncertainty relation to predict the measurement outcomes, it seems obviously clear that to protect entanglement from environmental noise. Many effort have been
done to achieve this purpose, such as dynamical decoupling \cite{VIOLA,VIOLA2,ZANARDI}, decoherence free subspaces \cite{LIDAR,XU,Feng}, quantum error correction code \cite{Calderbanka,Stean,Knill}, environment-assisted error correction scheme \cite{ZHAO} and quantum Zeno dynamics \cite{Facchi,Paz-Silva}. In litureture it is mentioned that the weak measurements and quantum measurement reversals can  protect the single qubit system from decoherence \cite{Sun,Korotkov,Xiao}. This important issue has also been developed on two-qubit  systems for protecting the entanglement from decoherence\cite{Sun1,Kim,Man,Li1}. In this work, the entropic uncertainty lower bound (EULB) in the presence of environmental noise is investigated. we consider the generalized amplitude damping channel and depolorizing channel as environmental noises. As expected, due to environmental effect the increasing of the (EULB) is inevitable. Here, we control the (EULB)  by using weak measurements and measurement reversals.  we will show that the (EULB) can be reduced to an optimal value by performing weak measurements, measurement reversal and regulating measurement parameters. This work is organized as follows: In Sec. \ref{2} We will review the concept of  generalized amplitude damping channel and depolarizing channel respectively. In Sec.\ref{3} we show how to use weak measurement to control the (EULB) in the presence of environmental noise .  We will examine a few examples In Sec. \ref{4}. The manuscript closes with conclusion and outlook in Sec.\ref{5}

\section{environmental noises}\label{2}
\subsection{Generalized amplitude damping channel}
When a system interacts with an environment at zero temperature, its evolution can be described by an amplitude damping (AD) channel  as follows 
\begin{equation}
\rho(t)=\sum_{i=0}^{1} E_{i}\rho(0) E_{i}^{\dag},
\end{equation}
where
\begin{equation}
 E_0 =  \begin{pmatrix} 1 & 0\\ 0 & \sqrt{1-p} \end{pmatrix}, \quad E_1 =  \begin{pmatrix} 0 & \sqrt{p}\\ 0 & 0 \end{pmatrix}, \nonumber \\
\end{equation}
are Kraus operators and $p$ represents the probability of transition from excited $\vert 1 \rangle $ to ground  state$\vert 0 \rangle$. At zero temperature, the only transition is the transition from a high energy level to a low energy level. It should be noted that having an environment at zero temperatures is not possible in practice. When the temperature of the environment is non-zero, the conditions are completely different. In this situation, in addition to losing excitation, the system can obtain excitation as a result of interaction with the environment. Such an interference with the environment can be described by the generalized amplitude damping (GAD) channel. By considering $r$ as a probability of losing excitation and $1-r$ as the probability of obtaining 
the excitation,  the Kraus operators of the (GAD) channel for two dimensional quantum systems are given by
\begin{align}
 E_0 &= \sqrt{r} \begin{pmatrix} 1 & 0\\ 0 & \sqrt{1-p} \end{pmatrix}, \nonumber \\
 E_1 &= \sqrt{r} \begin{pmatrix} 0 & \sqrt{p}\\ 0 & 0 \end{pmatrix}, \nonumber \\
 E_2 &= \sqrt{1-r} \begin{pmatrix} \sqrt{1-p} & 0\\ 0 & 1 \end{pmatrix}, \nonumber \\
 E_3 &= \sqrt{1-r} \begin{pmatrix} 0 & 0\\ \sqrt{p} & 0 \end{pmatrix}.
\end{align}
It is worth noting  tha, in the amount of $r=1$ (GAD) channel  is the same as (AD) channel. We consider a quantum bipartite system $AB$, such that each separate part interacts   with the environment at non-zero temperatures independently. The evolution of this quantum system in the Kraus representation  is given by
\begin{equation}
\rho_{AB}(t)= \sum_{i=0}^{5}(E_i \otimes E_{j})\rho_{AB}(0)(E_i \otimes E_{j})^{\dag},
\end{equation}
where $\rho_{AB}(0)$ is the initial state of the two-qubit system.

\begin{figure}[t]
\includegraphics[scale=1]{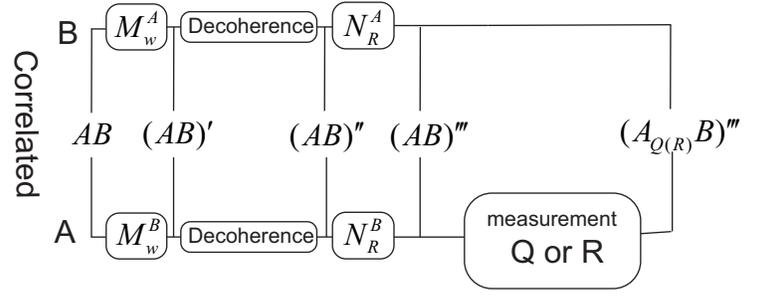}
\caption{ The scheme for Controlling (EULB)  using weak measurement and measurement reversal.}
\label{Fig1}
\end{figure} 
\subsection{Depolarizing channel}
The depolarizing  channel is a channel that depolarizes the state with  probability $r$  and leaves the state of the system unchanged with probability $1-r$. Depolarizing channel converts the state of the single-qubit systems to a completely mixed state. The Kraus operators of the depolarizing channel for two dimensional quantum systems are given by
\begin{align}
 F_0 &= \sqrt{1-r} \begin{pmatrix} 1 & 0\\ 0 & 1 \end{pmatrix},\quad F_1= \sqrt{\frac{r}{3}} \begin{pmatrix} 0 & 1\\1 & 0 \end{pmatrix}, \nonumber \\
 \nonumber \\
 F_2 &= \sqrt{\frac{r}{3}} \begin{pmatrix} 0 & -i \\ i & 0 \end{pmatrix},\quad F_3 = \sqrt{\frac{r}{3}} \begin{pmatrix} 1 & 0\\ 0 & -1 \end{pmatrix}, \nonumber \\
\end{align}
such that $r$ is the parameter of the depolarizing channel. Let us  consider a quantum bipartite system $AB$, such that each separate part interacts   with the depolarizing channel  independently. The evolution of this quantum system in the Kraus representation  is given by
\begin{equation}
\rho_{AB}(t)= \sum_{i=0}^{5}(F_i \otimes F_{j})\rho_{AB}(0)(F_i \otimes F_{j})^{\dag},
\end{equation}
where $\rho_{AB}(0)$ is the initial state of the two-qubit system.
\section{weak measurement and measurement reversal}\label{3}
Weak measurements are obtained by generalizing Von Neumann measurements. They are related to the positive
operator valued measure(POVM)\cite{Xiao1}. In general, a weak measurement operator  for a single-qubit systems is given by
\begin{equation}
M=   \left(   {\begin{array}{cc}
   1 & 0 \\
   0 & m \\
  \end{array} } \right) ,
\end{equation}

where $m \in [0,\infty )$. When  $m = 0$, the weak measurement is the same as projective measurement. For $0<m<1$, the weak measurement is a measurement which partially project the state 
on the ground state and for the case of $1<M<\infty$ the weak measurement represents  a measurement which partially project the state  on the excited state. The measurement reversal operator for single-qubit systems is given by 
\begin{equation}
N=   \left(   {\begin{array}{cc}
   n & 0 \\
   0 & 1 \\
  \end{array} } \right),
\end{equation}
where $n \in [0,\infty)$.

\section{Model}\label{4}
The fundamental method to control the (EULB) in the presence of environmental noise is illustrated in Fig. \ref{Fig1}. In Fig. \ref{Fig1} $M_w^A$($M_w^B$) represents weak measurement on subsystem $A$($B$) and $N_R^A$($N_R^B$) represents measurement reversal on subsystem $A$($B$). Bob prepares a correlated bipartite  quantum state $\rho_{AB}$ and sends part $A$ to Alice and holds the second part $B$ as a particle memory.  Then Alice and Bob will  reach an agreement for measuring  the two observable $\hat{Q}$ and $\hat{R}$ by Alice on her particle. Before the effect of the decoherence, the weak measurements $M_{AB}=M_w^A \otimes M_w^{B}$ is performed on quantum system 
\begin{equation}
M_{AB}=   \left(   {\begin{array}{cc}
   1 & 0 \\
   0 & m_{1} \\
  \end{array} } \right) \otimes \left(   {\begin{array}{cc}
   1 & 0 \\
   0 & m_{2} \\
  \end{array} } \right) .
\end{equation}
 The post weak measurement state of the system is given by
\begin{equation}
\rho_{AB}^\prime=\frac{M_{AB} \rho_{AB}  M_{AB}^{\dag}}{tr(\rho_{AB}M_{AB}^{\dag}M_{AB})}.
\end{equation}
In second step, quantum system $AB$  is affected by environment. The dynamics of such a system can be described by
\begin{equation}
\rho_{AB}^{\prime\prime}=\sum_{i,j=1}^{n}(K_{i}\otimes K_{j})\rho_{AB}^\prime(K_{i}\otimes K_{j})^{\dag},
\end{equation}
where $K_{i(j)}$'s are Kraus operators. Next in third step, the reversal measurement $N_{AB}=N_R^A \otimes N_R^B$ is performed on quantum system   
\begin{equation}
N_{AB}=   \left(   {\begin{array}{cc}
   n_1 & 0 \\
   0 & 1 \\
  \end{array} } \right) \otimes \left(   {\begin{array}{cc}
   n_2 & 0 \\
   0 & 1 \\
  \end{array} } \right),
\end{equation}
then the state of the system becomes
\begin{equation}
\rho_{AB}^{\prime\prime\prime}=\frac{ N_{AB} \rho_{AB}^{\prime\prime} N_{AB}^{\dag}}{tr(\rho_{AB}N_{AB}^{\dag}N_{AB})}.
\end{equation}
As the last step, According to the preliminary agreement between Alice and Bob, Alice performs a measurement on her particle. The post measurement state is given by
\begin{equation}
\rho_{OB}^{\prime\prime\prime}=\sum_{i} (\vert o_i \rangle \langle o_i \vert \otimes \mathcal{I})\rho_{AB}^{\prime\prime\prime} (\vert o_i \rangle \langle o_i \vert\otimes \mathcal{I}),
\end{equation} 
where $\lbrace \vert o_i \rangle \rbrace$’s are the eigenstates of the observable $\hat{O} \in \lbrace \hat{Q}, \hat{R} \rbrace$, and $\mathcal{I}$ is the identity operator. Thus the (EULB) is given by  $\log_{2}1/c + S(\rho_{A\vert B}^{\prime\prime\prime})$.
\section{Examples}\label{5}
\subsection{Bell diagonal state}
As a first example let us consider two-qubit Bell diagonal state  with the maximally mixed marginal states as an initial state which is shared between Alice and Bob. This state can be
written as
\begin{equation}
\rho_{AB}=\frac{1}{4}(\mathcal{I}\otimes \mathcal{I} + \sum_{i,j=1}^{3} w_{ij}\sigma_i \otimes \sigma_j),
\end{equation}
 where $\sigma_{i}$($i=1,2,3$) are the Pauli matrices.  By utilizing the singular value decomposition theorem, the matrix $W=\lbrace w_{ij}\rbrace$  can be diagonalized by a local unitary transformation, thus the Bell-diagonal states can be written as
 \begin{equation}\label{bell}
\rho_{AB}=\frac{1}{4}(\mathcal{I}\otimes \mathcal{I} + \sum_{i=1}^{3} c_i \sigma_i \otimes \sigma_i),
\end{equation}
where $\sigma_i$ ($i=1,2,3$) are Pauli matrices. This density matrix is positive if $\vec{c}=(c_1,c_2,c_3)$ belongs to a tetrahedron defined by the set of vertices $(-1,-1,-1)$,$(-1,1,1)$,$(1,-1,1)$ and $(1,1,-1)$. Here we consider the case which $c_{1}=1-2p$ and $c_{2}=c_{3}=-p$, with $0 \geq p \geq 1$. Thus the state in Eq.\ref{bell} can be written as  
\begin{equation}
\rho_{AB}=p \vert \Psi^{-} \rangle \langle  \Psi^{-} \vert + \frac{1-2p}{2}( \vert \Psi^{+} \rangle \langle  \Psi^{+} \vert +  \vert \Phi^{+} \rangle \langle  \Phi^{+} \vert).
\end{equation}
By following the process outlined in section \ref{4}, we can find  $\rho_{AB}^{\prime\prime\prime}$.
 Then, we check how the (EULB) behaves under the decoherence, weak measurement and measurement reversal. We obtained numerical results for the (EULB) with and without utilizing the weak measurement and  measurement reversal protocol. 
We select the weak measurement parameters in such a way that the weak measurement on the second part $B$ does not exist $m_2=1$ and we let $m_1=m$. One can get the minimum value of the (EULB) for an optimal  value of $m$.  In this work, we use the genetic algorithm to obtain the optimal value of (EULB).

\begin{figure}[t]
\includegraphics[scale=0.5]{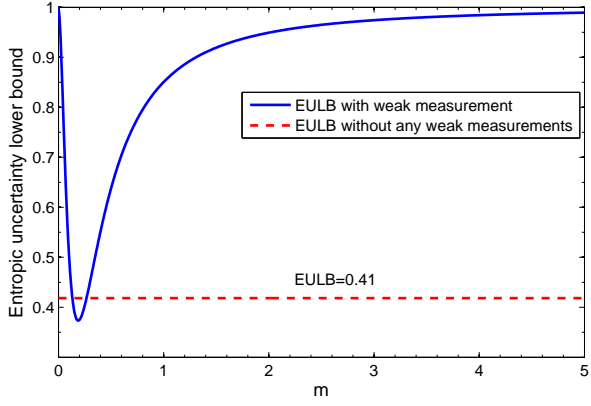}
\caption{(Color online) (EULB) as a function of the
weak measurement parameter $m$ with the (GAD) channel  parameters ($p_1=0.9, r_1=0.1, p_2=0.9, r_2=0.4$) and initial Bell diagonal state parameter $p=0$ (blue solid line) (EULB) without weak measurement and measurement reversal (red dashed line).}
\label{Fig2}
\end{figure}

\begin{figure}[t]
\includegraphics[scale=0.5]{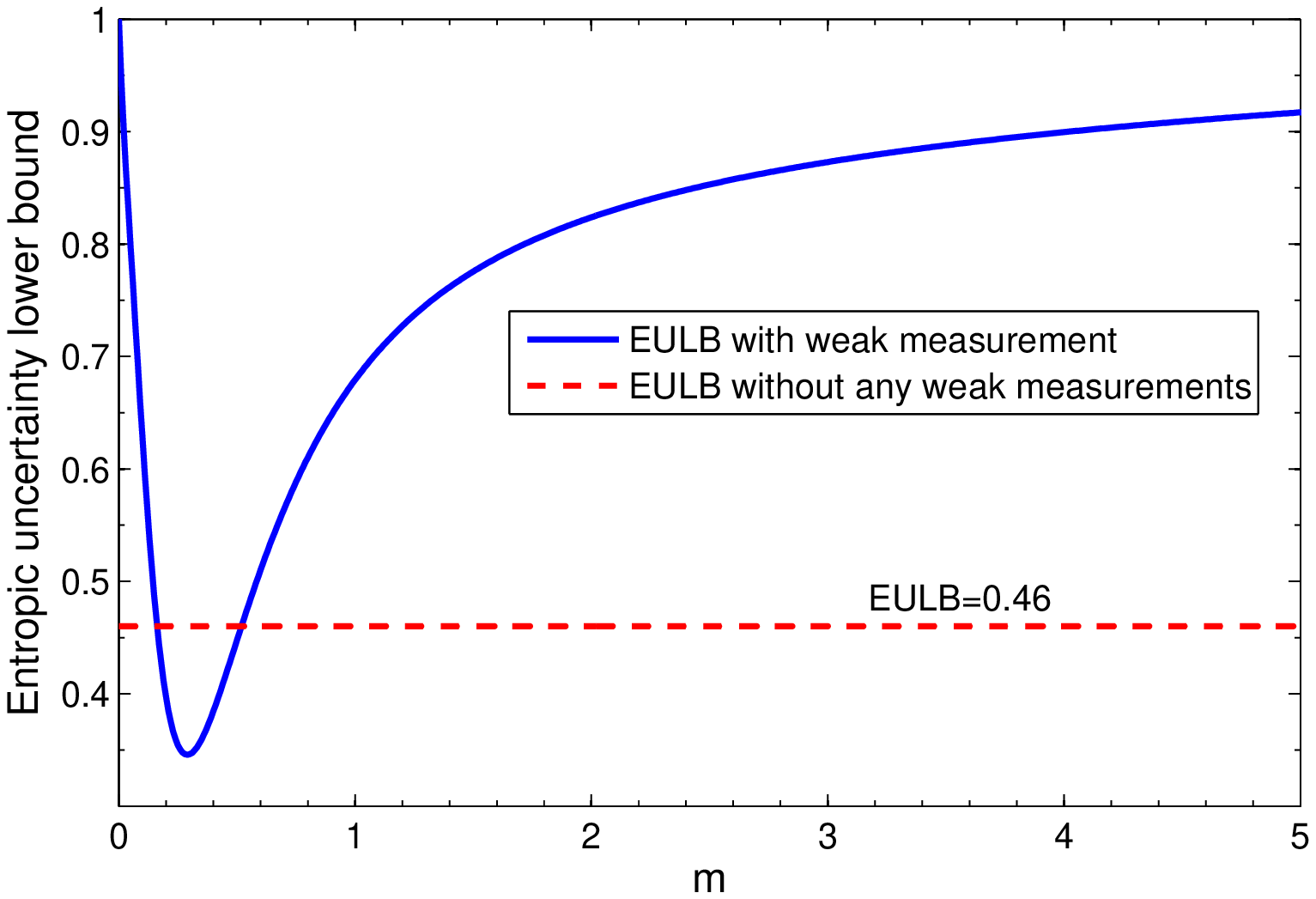}
\caption{(Color online) (EULB) as a function of the
weak measurement parameter $m$ with the (GAD) channel  parameters ($p_1=0.9, r_1=0.9, p_2=0.4, r_2=0.9$) and initial Bell diagonal state parameter $p=0.2$ (blue solid line) (EULB) without weak measurement and measurement reversal (red dashed line). }
\label{Fig3}
\end{figure}

\begin{figure}[t]
\includegraphics[scale=0.5]{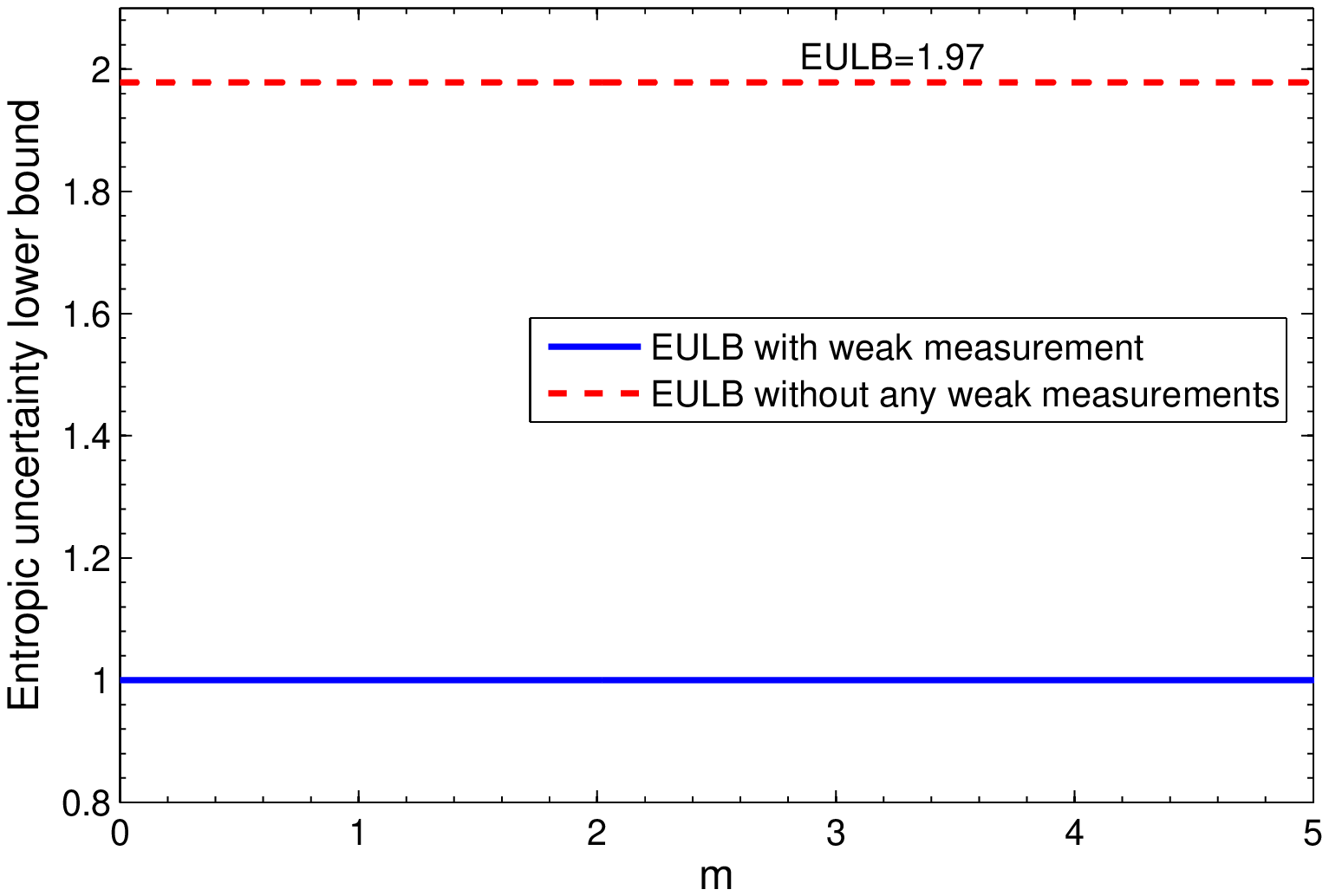}
\caption{(Color online) (EULB) as a function of the
weak measurement parameter $m$ with the depolarizing channel  parameters ($ r_1=0.9, r_2=0.9$) and initial Bell diagonal state parameter $p=0$ (blue solid line) (EULB) without weak measurement and measurement reversal (red dashed line).}
\label{Fig4}
\end{figure}

\begin{figure}[t]
\includegraphics[scale=0.5]{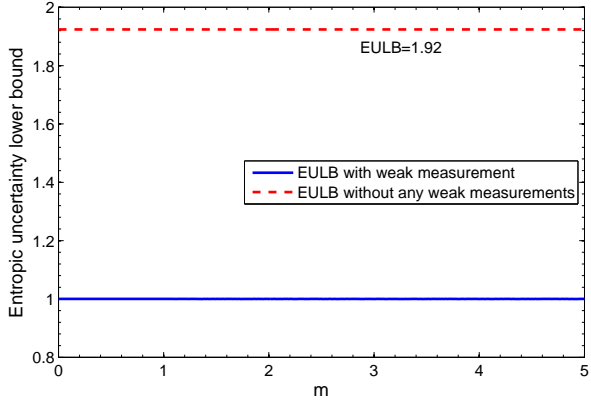}
\caption{(Color online) (EULB) as a function of the
weak measurement parameter $m$ with the depolarizing channel  parameters ($ r_1=0.9, r_2=0.1$) and initial Bell diagonal state parameter $p=1$ (blue solid line) (EULB) without weak measurement and measurement reversal (red dashed line).}
\label{Fig5}
\end{figure}

In order to illustrate the effects of the weak measurement and measurement reversal process, in Figs. \ref{Fig2}, \ref{Fig3}, \ref{Fig4}, and \ref{Fig5}  we plot (EULB)
in terms of weak measurement parameter $m$ for various initial states under generalized amplitude damping and depolarizing  channel. 
In Figs.\ref{Fig2} and \ref{Fig3} we plot (EULB) for two various initial state under the generalized amplitude damping channel characterized by decoherence parameters
($p_1$,$r_1$) and ($p_2$,$r_2$) for first and second part of the quantum state respectively. 
In Fig. \ref{Fig2} we use  the decoherence parameters ($p_1=0.9,r_1=0.1$), ($p_2=0.9,r_2=0.4$) and initial state parameter $p=0$. Through the optimization process we find that (EULB) reaches to its minimum value $0.37$ for ($m=0.18, n_1=0.18, n_2=0.81$). It is necessary to mention this point that without any weak measurements and measurement reversal the (EULB) is equal to $0.41$. In Fig. \ref{Fig3} the initial state parameter is $p=0.9$ and we use  the decoherence parameters ($p_1=0.1,r_1=0.9$), ($p_2=0.4,r_2=0.9$). In Fig \ref{Fig3} the minimum value of (EULB) is $0.34$, which is obtained for ($m=0.28, n_1=0.63, n_2=1.17$). It is worth noting that without any weak measurements and measurement reversal the (EULB) is equal to $0.46$. 
\begin{figure}[t]
\includegraphics[scale=0.5]{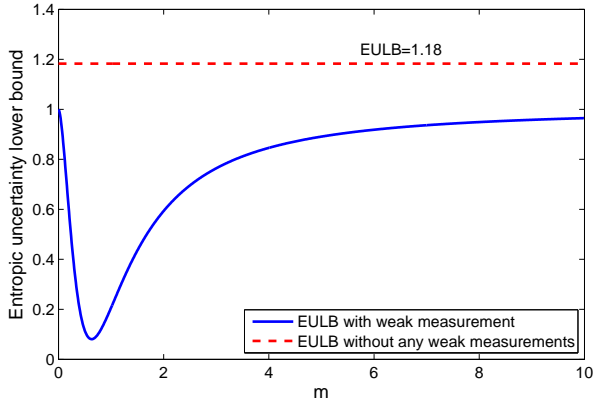}
\caption{(Color online) (EULB) as a function of the
weak measurement parameter $m$ with the (GAD) channel  parameters ($p_1=0.1, r_1=0.1, p_2=0.4, r_2=0.4$) and initial two qubit $X$ state parameter $p=0.5$ (blue solid line) (EULB) without weak measurement and measurement reversal (red dashed line).}
\label{Fig6}
\end{figure} 
\begin{figure}[t]
\includegraphics[scale=0.5]{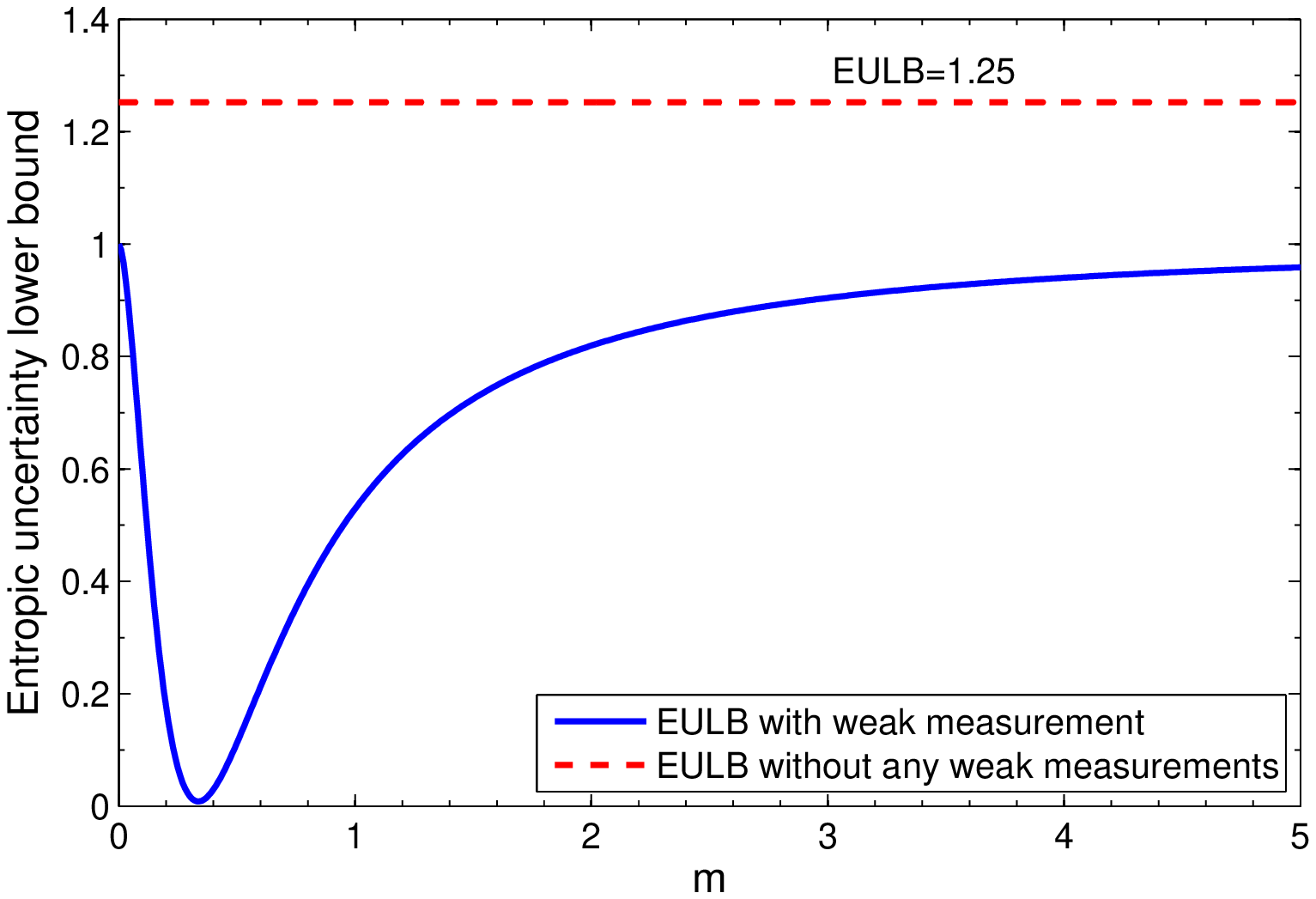}
\caption{(Color online) (EULB) as a function of the
weak measurement parameter $m$ with the (GAD) channel  parameters ($p_1=0.1, r_1=0.1, p_2=0.2, r_2=0.2$) and initial two qubit $X$ state parameter $p=0.2$ (blue solid line) (EULB) without weak measurement and measurement reversal (red dashed line).}
\label{Fig7}
\end{figure} 
\begin{figure}[t]
\includegraphics[scale=0.5]{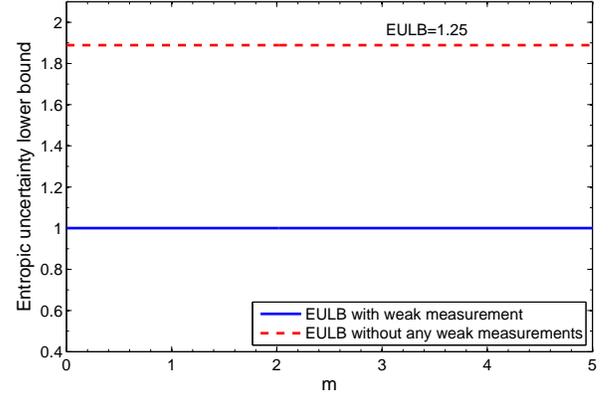}
\caption{(Color online) (EULB) as a function of the
weak measurement parameter $m$ with the depolarizing channel  parameters ($ r_1=0.4, r_2=0.1$) and initial two qubit  state parameter $p=0.5$ (blue solid line) (EULB) without weak measurement and measurement reversal (red dashed line).}
\label{Fig8}
\end{figure} 
Now we focus our attention on the dynamics of the (EULB) in terms of the  weak measurement parameter $m$ for two various Bell diagonal initial-state parameters under the depolarizing channel. In Fig. \ref{Fig4} the initial state parameter is $p=0$ and we use  the depolarizing channel parameters ($r_1=0.1,r_2=0.9$). From Fig. \ref{Fig4} we find that (EULB) reaches to its minimum value $1$ for ($m=1.16, n_1=2.9 \times 10^{-7}, n_2=0.09$). In this situation, without existing any weak measurements and measurement reversal the (EULB) is equal to $1.97$. In Fig. \ref{Fig5} we use  the depolarizing channel parameters ($r_1=0.9,r_2=0.1$) and initial state parameter $p=1$. Here the minimum value of (EULB) is $1$, which is obtained for ($m=0.35, n_1=5.7\times 10^{-7}, n_2=0.1$). When there is no weak measurement and measurement reversal the (EULB) is equal to $1.92$.  By comparing the results for two conditions, the existence and absence of any weak measurement. One  can easily conclude that if the weak measurement  and measurement reversal do not exist i.e. ($m_1=n_1=n_2=1$), the decoherence part of the model in Fig. \ref{Fig1} leads the (EULB) increases rapidly. Therefore, we have shown that the weak measurements and
measurement reversal have reduced the (EULB) under the decoherence.
\subsection{Two-qubit X states
}
In this section, we consider the special class of two-qubit states that are called $X$-states as an initial state.
\begin{equation}
\rho_{AB}=p \vert \psi^{+}\rangle \langle \psi^{+} \vert + (1-p)\vert 11 \rangle \langle 11 \vert,
\end{equation}
where $ \vert \psi ^{+} \rangle = \frac{1}{\sqrt{2}}(\vert 01 \rangle + \vert 10 \rangle)$ is a maximally entangled state and $0 \leq p \leq 1$. In Fig. \ref{Fig6} the initial state parameter is $p=0.5$ and we use  the (GAD) channel parameters ($p_1=0.1,r_1=0.1$), ($p_2=0.4,r_2=0.4$).  As can be seen from Fig. \ref{Fig6},  (EULB) reaches to its minimum value $0.08$ for ($m=0.63, n_1=9619.9, n_2=15365.6$). In this situation, without existing any weak measurements and measurement reversal the (EULB) is equal to $1.18$.  Fig. \ref{Fig7}  shows the (EULB) for initial state parameter $p=0.2$ and the (GAD) channel parameters ($p_1=0.1,r_1=0.1$), ($p_2=0.2,r_2=0.2$).  As can be seen from Fig. \ref{Fig7},  (EULB) reaches to its minimum value $0.008$ for ($m=0.33, n_1=658.7, n_2=2006.9$). In this situation,  without existing any weak measurements and measurement reversal the (EULB) is equal to $1.25$.  

We are now focusing on the evolution  of the (EULB) in terms of the  weak measurement parameter $m$ for two various initial $X$-state  under the depolarizing channel.

\begin{figure}[t]
\includegraphics[scale=0.5]{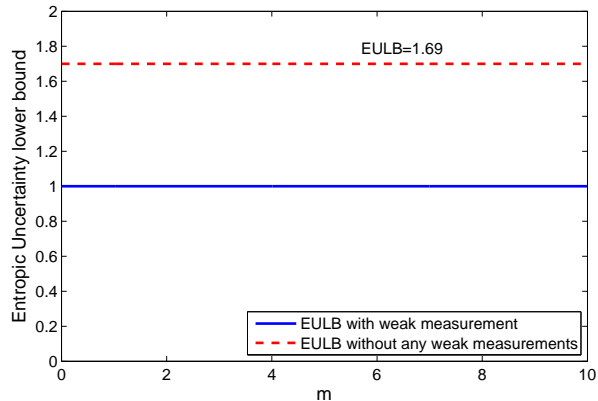}
\caption{(Color online) (EULB) as a function of the
weak measurement parameter $m$ with the (GAD) channel  parameters ($p_1=0.1, r_1=0.1, p_2=0.2, r_2=0.2$) and initial two qubit $X$ state parameter $p=0.2$ (blue solid line) (EULB) without weak measurement and measurement reversal (red dashed line).}
\label{Fig9}
\end{figure}

 In Fig. \ref{Fig8}, we consider the dynamics of the (EULB) in terms of the  weak measurement parameter $m$ for initial two qubit $X$- state parameters $p=0.5$ under the depolarizing channel with parameters ($r_1=0.4, r_2=0.1$). As can be seen from Fig. \ref{Fig8},  (EULB) reaches to its minimum value $1$ for ($m=1.5, n_1=5.1\times 10^{-9}, n_2=0.54$). In this situation, without existing  any weak measurements and measurement reversal the (EULB) is equal to $1.88$.  In Fig. \ref{Fig9} we use  the depolarizing channel parameters ($r_1=0.4,r_2=0.1$) and initial two qubit $X$-state state  with parameter $p=1$. As can be seen from Fig. \ref{Fig9},  (EULB) reaches to its minimum value $1$ for ($m=1.15, n_1=2.2\times 10^{-6}, n_2=0.32$). In this situation, without existing  any weak measurements and measurement reversal the (EULB) is equal to $1.69$.
By comparing the results for two conditions, the existence and absence of any weak measurement. One  can easily conclude that if the weak measurement  and measurement reversal do not exist i.e. ($m_1=n_1=n_2=1$), 
then (EULB) increases rapidly. Thus we find that the weak measurement and measurement reversal can decrease the (EULB) under the decoherence.
\section{CONCLUSION AND OUTLOOK}\label{6}
In this work we studied the effects of the environment on the entropic uncertainty lower bound in the presence of weak measurement and measurement reversal. First the weak measurement is performed on bipartite quantum system $AB$, then the decoherence affects on each part of the system independently, and at last the measurement reversal is performed on decohered system. Here we considered the generalized amplitude damping channel and depolarizing channel as the environmental noises. We consider two various initial state, two qubit Bell diagonal and two qubit $X$-state and different  parameters for (GAD) and depolarizing channel. In this work, we showed that the weak measurement and measurement reversal is an appropriate tool for protecting (EULB) from enhancing under the decoherence. In this manuscript we observed that  by regulating weak measurement and measurement reversal parameters the (EULB) dropped to an optimal minimum value.

Based on the results shown in the literature, weak measurement and measurement reversal enhance the quantum correlation in bipartite quantum systems \cite{Xiao1,Doustimotlagh,Basit,Sun,Guo}. Thus, the result we have taken here is quite reasonable, Because by enhancing the correlation between the memory particle (which is in Bob's possession) with the measured particle (which is in Bob's possession) the entropic uncertainty lower bound decreases to optimal value.

\end{document}